\documentclass[twocolumn,showpacs,preprintnumbers,amsmath,amssymb]{revtex4}
\usepackage{amsmath,amsfonts,diagbox,latexsym,amssymb,graphics,epsfig,subfigure,color,makeidx}
\usepackage{xcolor}
\usepackage[colorlinks,
            linkcolor=blue,
            anchorcolor=blue,
            citecolor=green
            ]{hyperref}

\begin{document}

\title{Dynamics of scalar hair with self-interaction around Schwarzchild black hole}

\author{Yu-Peng Zhang\footnote{zyp@lzu.edu.cn},
	Yong-Qiang Wang\footnote{yqwang@lzu.edu.cn},
	Shao-Wen Wei\footnote{weishw@lzu.edu.cn},
	Yu-Xiao Liu\footnote{liuyx@lzu.edu.cn, corresponding author}}

\affiliation{Lanzhou Center for Theoretical Physics, Key Laboratory of Theoretical Physics of Gansu Province, School of Physical Science and Technology, Lanzhou University, Lanzhou 730000, China\\
	Institute of Theoretical Physics \& Research Center of Gravitation, Lanzhou University, Lanzhou 730000, China\\
}

\begin{abstract}
In this paper, we study the dynamics of scalar hair around a Schwarzchild black hole. The scalar hair is sourced by the Gauss-Bonnet invariant with a linear coupling. We work perturbatively in the coupling constant and ignore the back-reaction of the scalar hair and the Gauss-Bonnet invariant. We evolve the scalar field in the background of a Schwarzchild black hole and study the dynamical formation of scalar hair with different self-interactions. We integrate the energy and compute the energy flux of the scalar hair in terms of the canonical energy-momentum tensor and give the corresponding dependence on the self-interactions. Our results allow us to estimate the radiation and the condensation of scalar field with different self-interactions, these would improve our understanding for the dynamical scalarization of black holes and the possible configuration of the scalar hair in scalar Gauss-Bonnet gravity.

\end{abstract}

\maketitle

\section{Introduction}\label{scheme1}
The recent observations of gravitational waves (GWs) \cite{Abbott2016a} and black hole shadow \cite{ETH2019} opened the era to study the properties of strong gravity systems, which further support the black hole hypothesis and promise to reveal new insights into the structure of black holes. In general relativity (GR), the structure of a black hole can be determined by the mass $M$, charge $Q$, and spin angular momentum $J$ \cite{Hawking:1971vc,Wald:1971iw,Carter:1971zc,Bekenstein:1995un}. In principle, one can obtain the serious constraint on the parameters of a black hole in terms of the observations of GWs or the shadow information if these data are accurate enough. However, besides GR, other alternative theories of gravity have also been focused and the corresponding history is almost as old as that of GR \cite{Will:2014kxa,Berti:2015itd} and various black holes are proposed that deviate from the ones in GR.

As a fundamental matter field, the scalar field provides rich of models for understanding some unknown phenomena in our universe, such as the dark matter \cite{Marsh:2015xka}, the inflation of cosmology \cite{Bezrukov:2007,Burgess2009,Burgess2010,Giudice2011,Lerner2010,Lyth1999}. The scalar field is also very useful to understand the formation of a black hole,  like the critical behavior in the gravitational collapse \cite{Choptuik1993,Goncalves:1997qp,Healy:2013xia,Kidder2019}. At the same time, the scalar field can also be considered as a bridge between GR and modified gravity theories. For example, the scalar field can be regarded as the effective degree of freedom of scalar-tensor and $f(R)$ theories \cite{Wagoner1970,Felice2010,Sotiriou2010}, for which the classical scalar field plays the role of the modification of GR. The observations of GWs from the black hole mergers make sure that black holes can be considered as the perfect probes for testing various alternative theories of gravity, and one can also study the structures of black holes by observing the imprint of the scalar field.

Note that, the no-hair theorems \cite{Hawking:1972qk,Bekenstein:1995un} proved that minimally coupled, potentially self-interacting scalar fields all have the trivial configurations around the stationary and asymptotically flat black holes. However, when the nonminimally coupling mechanism between the scalar field and Gauss-Bonnet (GB) term is considered \cite{Sotiriou:2013qea,Sotiriou:2014pfa,Maselli:2015yva}, the no-hair theorems are violated and the nontrivial scalar configurations can form around black holes. The existence of the nontrivial scalar configurations around a black hole uncovers the deviation from the predictions of GR and provides the experimental insight into various alternative theories of gravity. Since the scalar GB gravity theory \cite{Sotiriou:2013qea,Sotiriou:2014pfa,Maselli:2015yva} was proposed, this theory has attracted a lot of attention and the spontaneous scalarization of a black hole has been extensively studied in recent years \cite{Doneva:2017bvd,Silva:2017uqg,Andreou:2019ikc,Hod:2019pmb,Peng:2019snv,Ripley:2019aqj,Hod:2019vut,Dima:2020yac,Liu:2020yqa,Guo:2020sdu,Herdeiro:2020wei,East:2021bqk}. Besides the scalarization induced by the GB term, other scalarization mechanisms of black holes were also extensively studied \cite{Gregory:1992kr,Tang:2020sjs,Wang:2020ohb,Guo:2021zed,Yao:2021zid,Zhang:2021ybj,Zhang:2021nnn}.

In the appropriate limit, several static scalar hairy black hole solutions have been proposed \cite{Silva:2017uqg,Andreou:2019ikc,Hod:2019pmb,Peng:2019snv,Ripley:2019aqj,Hod:2019vut,Dima:2020yac,Liu:2020yqa,Guo:2020sdu,Herdeiro:2020wei,East:2021bqk}. However, the dynamical formation of scalar hair has not been disclosed. To study the dynamical formation of scalar hair, one should evolve the spacetime and scalar field in the scalar GB gravity \cite{Julie:2020vov,Witek:2020uzz,East:2020hgw,Doneva:2021tvn}. Note that, it is difficult to realize the nonlinear evolution of a hairy black hole in the scalar GB gravity due to the complexity of the field equations. For simplicity, one can also figure out the part of the dynamics of the scalar hair under the linear approximation. Reference \cite{Benkel:2016kcq} firstly investigated the dynamical formation of the scalar hair around a Schwarzchild black hole under a linear perturbation with the coupling constant but
ignoring the back-reaction of the scalar hair and the GB invariant. The results showed that the evolution eventually settles to the known static hairy solutions in the appropriate limit, see the details in Ref. \cite{Benkel:2016rlz}. The dynamics of scalar hair in the rotating black hole and binary mergers were also studied in Ref. \cite{Witek:2018dmd}, and the scalar radiation and multi-pole waveforms of scalar field in the backgrounds of binary mergers were obtained. In addition, the dynamics of
scalar hairs around a Schwarzchild black hole both in GB
gravity and Chern-Simmons gravity were considered \cite{Doneva:2020nbb,Doneva:2021dcc,Doneva:2021dqn}.

{It has been shown that the scalar configurations will be different when the self-interactions are considered \cite{Gregory:1992kr,Doneva:2019vuh,Macedo:2019sem,Doneva:2020kfv}. For example, the number of the horizons of a black hole in massive dilaton gravity {depends on} the product of the dilaton mass and the black hole charge \cite{Gregory:1992kr}. Similarly, the corresponding dynamics of {the} massive scalar hair will also change. In previous Refs. \cite{Benkel:2016kcq,Benkel:2016rlz}, the dynamical formation of the massless scalar hair around a Schwarzchild black hole has been investigated. Compared with the massless scalar field, we consider a massive scalar {field with} the quartic self-interaction $\lambda \phi^4$, {where $\lambda$ is the coupling parameter}. For the massive scalar field its Compton wavelength $\lambda_\phi \sim \hbar c/m_\phi$ is related to the mass parameter $m_\phi$ \cite{Dolan:2007mj} and the corresponding dynamics will be different. For example, when the Compton wavelength of the massive boson field is comparable to or larger than the black hole horizon radius, there will exist {one or more} bound states around the black hole. If the black hole is spinning, a bound state can grow from a seed perturbation through the superradiance mechanism \cite{Dolan:2007mj,Witek:2012tr,East:2017mrj}, where the growth rate and the maximal extracted energy {also depends on} the mass parameter \cite{Dolan:2007mj,Witek:2012tr,East:2017mrj}. For the quartic interaction $\lambda \phi^4$, it will have a nontrivial repulsive effect when the parameter satisfies $\lambda>0$. Such repulsive effect is also related to the stability of the scalar hair. Reference \cite{ Macedo:2019sem} has shown that for a fixed mass parameter $m_\phi$, there is a threshold $\lambda_{\text{crit}}$ and when $\lambda<\lambda_{\text{crit}}$ the scalarized solutions are unstable.
	
Therefore, it is natural to expect that the growth rate, the maximal energy of the canonical part, {and the} dynamics of the scalar hair will be changed when the mass term and quartic self-interaction are considered. In this paper, we focus on the dynamics of a massive scalar field with the quartic self-interaction $\lambda \phi^4$ around a Schwarzchild black hole in the scalar GB gravity. We will study the dynamical formation of the scalar hair and show how the growth rate, the energy flux, and the maximal energy of the canonical part of the scalar hair are affected by the mass term and the quartic self-interaction.} We work perturbatively in the coupling constant and ignore the back-reaction of the scalar hair and the GB invariant. We adopt the same way in Ref. \cite{Benkel:2016kcq} to study the evolution of the scalar field.

Our paper is organized as follows. In Sec.~\ref{scheme1}, we briefly review the knowledge of the scalar GB gravity and give the corresponding equations of motion for the gravitational fields and scalar field. In Sec.~\ref{scheme2}, we give the numerical results and the corresponding analysis. Finally, a brief conclusion and outlook are given in Sec.~\ref{Conclusion}.

\section{Setup}{\label{scheme1}}
\subsection{Action and field equations}
In this paper, we study the dynamical formation of scalar hair around a Schwarzchild black hole in the scalar GB gravity. The action of the system is \cite{Benkel:2016rlz,Benkel:2016kcq}
\begin{eqnarray}
S &=&  \int d^4 x \sqrt{-g}
     \Bigg\{ \frac{R}{16\pi G} + \mu\bigg[-\frac{1}{2}\nabla_\mu\phi\nabla^\mu\phi \nonumber\nonumber\\
     && - V(\phi)+\eta f(\phi) \mathcal{G}\bigg]\Bigg\},
\label{action}
\end{eqnarray}
where the GB invariant $\mathcal{G}$ is defined as
\begin{equation}
\mathcal{G}={R^{\mu\nu\rho\sigma}R_{\mu\nu\rho\sigma}-4R^{\mu\nu} R_{\mu\nu}}+R^2.
\end{equation}
The scalar field is coupled with the GB invariant with $\eta f(\phi)\mathcal{G}$ and $\eta$ is the coupling constant. $V(\phi)$ is the scalar potential that describes the self-interactions of the scalar field. The Newton gravitational constant $G$, the speed of light $c$, and the Planck constant $\hbar$ are set to be unity ($G=c=\hbar=1$). The potential $V(\phi)$ is taken as the simplest form
\begin{equation}
V(\phi)=\frac{1}{2}m_\phi^2\phi^2 + \frac{1}{2}\lambda \phi^4,
\label{potentialofphi}
\end{equation}
where $m_\phi$ is the mass parameter of the scalar field the dimensionless parameter $\lambda$ is the coupling parameter of the self-interaction $\phi^4$. We define a dimensionless mass parameter $\bar{m}_\phi=m_\phi/\mathcal{M}$ with $\mathcal{M}$ a new mass scale. In this paper, we consider the following linear coupling
\begin{equation}
f(\phi)=\phi.
\label{effective-potential-r}
\end{equation}

Varying the action \eqref{action} with respect to the metric $g_{\mu\nu}$ and the scalar field $\phi$, we get the following fields equations
\begin{eqnarray}
G_{\mu\nu}+16\pi\mu\eta \mathcal{G}_{\mu\nu}^{\text{GB}}&=&8\pi\mu T_{\mu\nu}^{(\phi)},
\label{eom_gravity}\\
\square \phi - \frac{\partial V(\phi)}{\partial \phi}&=&\eta\mathcal{G}.
\label{eom_phi}
\end{eqnarray}
The energy-momentum tensor of the canonical scalar field is
\begin{equation}
T_{\mu\nu}^{(\phi)}=\nabla_\mu \phi \nabla_\nu \phi-\frac{1}{2}g_{\mu\nu}\left[\nabla^\mu \phi \nabla_\mu \phi + V(\phi)\right].
\label{tmunu_phi}
\end{equation}
The energy-momentum tensor contributed from the Gauss-Bonnet term is
\begin{eqnarray}
\mathcal{G}_{\mu\nu}^{\text{GB}}&=&-2R\nabla_{(\mu}\nabla_{\nu)}\phi-4R_{\mu\nu}\square\phi+4R_{\mu\alpha\nu\beta}\nabla^\alpha\nabla^\beta\phi\nonumber\\
& &+8R_{\alpha(\mu}\nabla^\alpha\nabla_{\nu)}\phi + 2g_{\mu\nu}\left(R\square\phi - 2R_{\alpha\beta}\nabla_\alpha\nabla_\beta\phi\right)\nonumber\\
&=&g_{\alpha(\mu}g_{\nu)\beta}\epsilon^{\alpha\gamma\delta\xi}\epsilon^{\varphi\beta\chi\theta}R_{\delta\xi\chi\theta}\nabla_{\gamma}\nabla_{\varphi}\phi.
\end{eqnarray}

We would like to study the evolution of the scalar field and its imprint on the spacetime of Schwarzchild black hole. However, equation \eqref{eom_gravity} is hard to evolve and we will consider the evolution of the scalar field and hair formation in the given spacetime background of a Schwarzchild black hole. In the decoupling limit $\mu\to 0$, the back-reaction of the scalar field on the metric can be ignored. Under this limit, the field equation \eqref{eom_gravity} reduces to the Einstein equation in vacuum and the equation of motion for the scalar field remains unaffected. Then we have
\begin{equation}
G_{\mu\nu}=0.
\label{eom_gravity_b}
\end{equation}

\subsection{Spacetime split and evolution of scalar field}
We will numerically evolve the system of Eqs. \eqref{eom_phi} and \eqref{eom_gravity_b}. The $3+1$ decomposition
\begin{equation}
ds^2=-\alpha^2dt^2+\gamma_{ij}(\beta^i dt+ dx^i)(\beta^j dt+ dx^j)\label{admmetric}
\end{equation}
is used to decompose the four-dimensional spacetime into a family of three-dimensional spacelike hypersurface ($\mathcal{\sum}_t, \gamma_{ij}$). Here, $\alpha(t,x^i)$ is the lapse function, $\beta^i$ is the shift vector, and $\gamma_{ij}$ is the spatial metric. The relation between the spacial metric $\gamma_{ij}$ and the spacetime metric $g_{\mu\nu}$ reads
\begin{equation}
\gamma^{\mu\nu}=g^{\mu\nu}+n^\mu n^\nu,
\end{equation}
where $n^{\mu}$ is the timelike normal vector and defined as
\begin{equation}
n^\mu=(\alpha^{-1},-\alpha^{-1}\beta^i).
\end{equation}
The Schwarzchild black hole solution can be obtained by using the ADM-York decomposition in numerical relativity \cite{York:1978gql} and the corresponding line element in isotropic coordinates is given by
\begin{eqnarray}
ds^2&=&-\left(\frac{1-\frac{M}{2r}}{1+\frac{M}{2r}}\right)dt^2+\left(1+\frac{M}{2r}\right)^4\eta_{ij}dx^idx^j\nonumber\\
&=&-\alpha^2dt^2+\psi^4\eta_{ij}dx^idx^j,
\end{eqnarray}
where $M$ is the bare mass of the black hole and $r$ is the isotropic, radial coordinate. We set $M=1$ in the following parts of this paper. One can take a transformation $\phi\to\eta\phi$ to drop out $\eta$ from Eq. \eqref{eom_phi} \cite{Benkel:2016rlz,Benkel:2016kcq} and we will just set the dimensionless coupling constant $\bar{\eta}=\eta M=1$ in the following parts.

The stable evolution of spacetime is realized by the Baumgarte and Shapiro \cite{Shapiro1999}, Shibata and Nakamura \cite{Shibata1995} (BSSN) forms. Using the $(3+1)$ metric (\ref{admmetric}) and the following conjugate momentum
\begin{equation}
\Pi=-\frac{1}{\alpha}(\partial_t\phi-\beta^i\partial_i\phi)
\end{equation}
of the scalar field, the equation of motion for the scalar field \eqref{eom_phi} becomes
\begin{eqnarray}
\partial_t\phi&=&\beta^i\partial_i\phi-\alpha\Pi,\\
\partial_t\Pi&=&\beta^k\partial_k\Pi - \alpha D^iD_i\phi -\gamma^{ij}D_i\alpha D_j\phi+ \alpha K\Pi\nonumber\nonumber\\
&&+\alpha\frac{\partial V(\phi)}{\partial \phi}+\alpha\eta\mathcal{G}.
\end{eqnarray}
Here, we are working on the decoupling limit and the GB term is only dependent on the background geometry. Reference \cite{Benkel:2016rlz} has shown that the different initial configurations of the scalar field do not affect the corresponding final state, and the evolution will eventually settle to the known static hairy solutions in the appropriate limit. Therefore, we set the initial data with a trivial scalar field configuration
\begin{equation}
\phi_0=\Pi_0=0.
\label{initial_data}
\end{equation}

Our simulations are based on the Maya numerical relativity code \cite{hhdl2007,bifd2007,Healy2009,Bode2010,Bode2011,Bode2012,Shcherbakov2012,Healy2012,kjjlpd2016}. The evolution of the spacetime for Maya code is based on the BSSN formulation of the Einstein equations \cite{Shapiro1999,Shibata1995}, and the moving puncture gauge condition \cite{Campanelli2005,Baker2006} is adopted.
Maya is compatible with the Einstein Toolkit \cite{EinsteinToolkit}. The Sommerfeld boundary conditions are adopted for our system to avoid the unphysical reflections. The BSSN formulation \cite{Shapiro1999,Shibata1995} and the moving puncture gauge condition \cite{Campanelli2005,Baker2006} ensure that the GB invariant $\mathcal{G}$ is regular everywhere.

\section{Numerical results}\label{scheme2}

In this paper, we focus on the dynamics of scalar hair with the self-interaction \eqref{potentialofphi}. {Due to the Compton wavelength $\lambda_\phi \sim \hbar c/m_\phi$ of the scalar field is determined by the mass parameter $m_\phi$ \cite{Dolan:2007mj}, we choose four values of the mass parameter $m_\phi$ and let the corresponding Compton {wavelength $\lambda_\phi$ be larger} than, comparable, or smaller than the horizon radius of the central black hole.} We still set the different values for the parameters $\lambda$ and evolve $16$ cases in total, see the details in Table \ref{values_of_parameters}. {We set the resolution of the innermost region to be $h=M/32$ let the radius of the finest level of our grid be $1.25M$, which ensures that the finest level can cover the central black hole.} The setup for the refinement boxes is
\begin{eqnarray}
\{(160,80,40,20,10,5,2.5,1.25)M\nonumber\}.
\end{eqnarray}

\begin{table}[!htb]
\begin{center}
\caption{Values of the parameters ($\bar{m}_{\phi}$, $\lambda$) considered in our model and the corresponding aliases.}
\begin{tabular}{ c| c  c  c  c}
\hline
\hline
\diagbox{$\bar{m}_{\phi}$}{$\lambda$} &~~0.0~~&~~10~~&~~100~~&~~1000~~\\
\hline
~~0.0~              &~~~sf\_1\_1~~~&~~~sf\_1\_2~~~&~~~sf\_1\_3~~~&~~~sf\_1\_4~~~\\
\hline
~~0.5~             &~~~sf\_2\_1~~~&~~~sf\_2\_2~~~&~~~sf\_2\_3~~~&~~~sf\_2\_4~~~\\
\hline
~~1.0~              &~~~sf\_3\_1~~~&~~~sf\_3\_2~~~&~~~sf\_3\_3~~~&~~~sf\_3\_4~~~\\
\hline
~~1.5~              &~~~sf\_4\_1~~~&~~~sf\_4\_2~~~&~~~sf\_4\_3~~~&~~~sf\_4\_4~~~\\
\hline
\hline
\end{tabular}
\label{values_of_parameters}
\end{center}
\end{table}

\begin{figure*}[!htb]
	\includegraphics[width=0.98\linewidth]{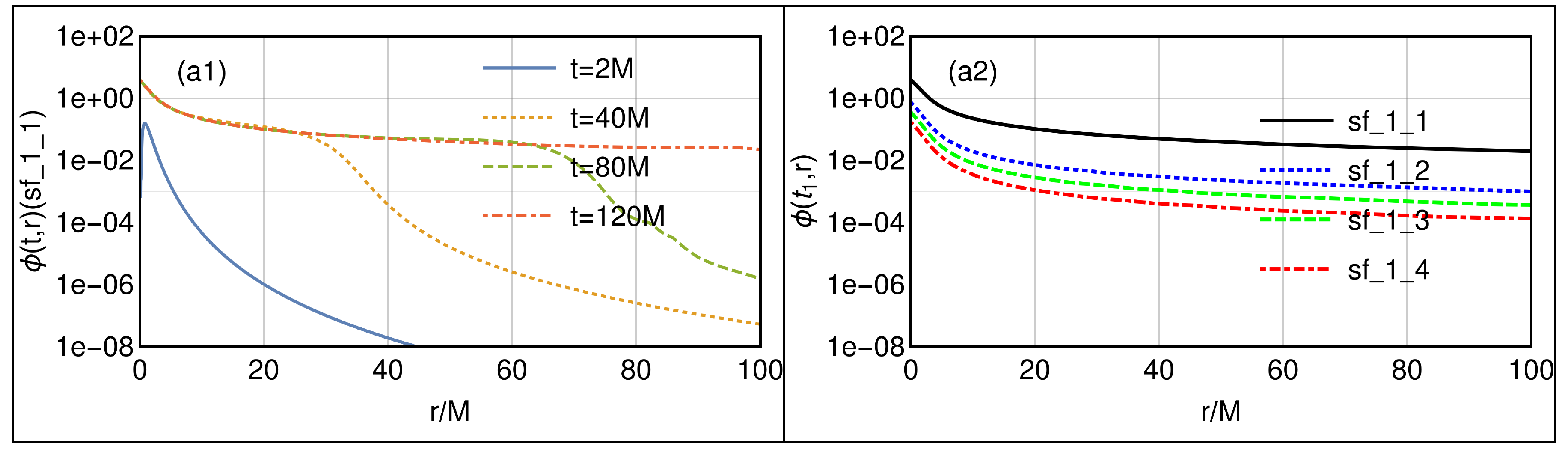}
	\includegraphics[width=0.98\linewidth]{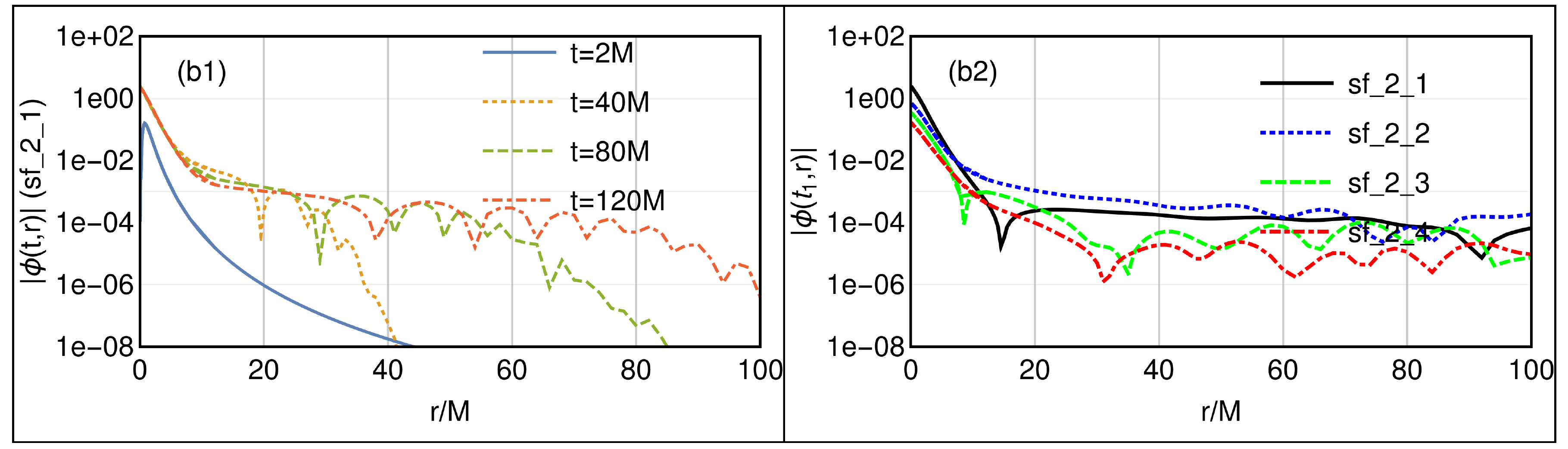}
	\includegraphics[width=0.98\linewidth]{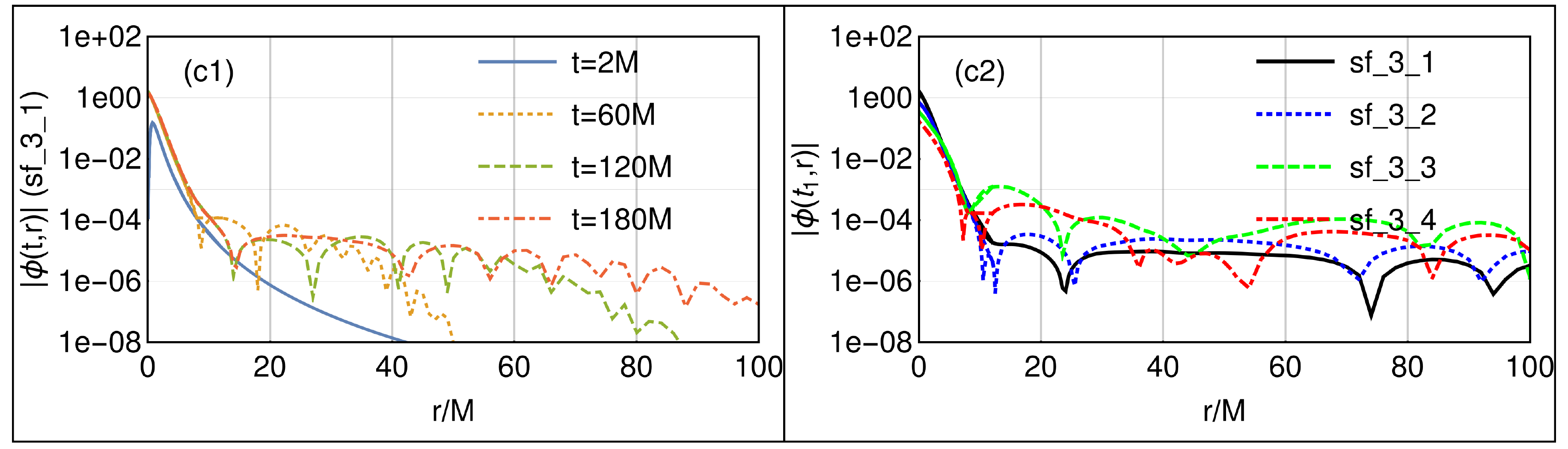}
	\includegraphics[width=0.98\linewidth]{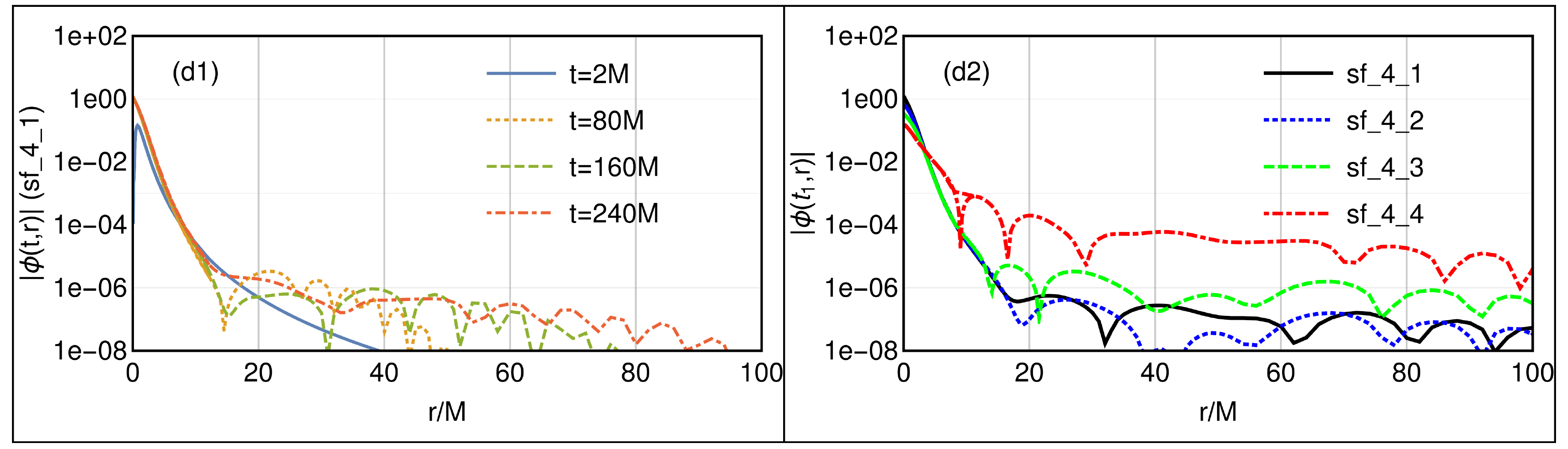}
	\caption{Radial profiles of the scalar field for the cases described in Table \ref{values_of_parameters}. The subfigures (a1), (b1), (c1), and (d1) describe the radial profiles of the scalar field at different instances of time for the cases of (sf\_1\_1, sf\_2\_1, sf\_3\_1, sf\_4\_1). The subfigures (a2), (b2), (c2), and (d2) describe radial profiles of scalar fields at late time $t_1=400M$.}
	\label{radial_profiles_of_sf}
\end{figure*}

We give the radial profiles of the scalar field at different instances during the evolution with the initial data \eqref{initial_data}.  Figure \ref{radial_profiles_of_sf} shows the radial profile of the scalar field for each case. It is found that the massless scalar field quickly relaxes to the static, hairy configurations at late time for the cases of (sf\_1\_1, sf\_1\_2, sf\_1\_3, sf\_1\_4). We observe that there are no oscillate behaviors and they are positive everywhere for the massless scalar field at late time. Comparing the corresponding static configurations at late time, we observe the values of the scalar field at late time decrease with the coupling parameter $\lambda$.

For the massive scalar field, we find that they will oscillate and then gradually decay to static, hairy scalar configurations. Comparing with the results in subfigures (b2), (c2), and (d2) in Fig. \ref{radial_profiles_of_sf}, we find that the value of the scalar field decreases with the mass of the scalar field. We project the scalar field by using the following spherical harmonics
\begin{equation}
\phi_{lm}(t,r)=\int{d\Omega \phi(t,r,\theta,\varphi)}Y^{*}_{lm}(\theta,\varphi),
\label{sphericalharmonic}
\end{equation}
and measure the values of $\phi_{lm}(t,r)$ at a fixed radius. The projection of the scalar field with spherical harmonics is realized with the help of the thorn Multipole. Note that only the $00$ mode of the scalar field exists because the scalar field is sourced by the spherical GB term in the background of the Schwarzchild black hole.

\begin{figure*}[!htb]
	\includegraphics[width=1.0\linewidth]{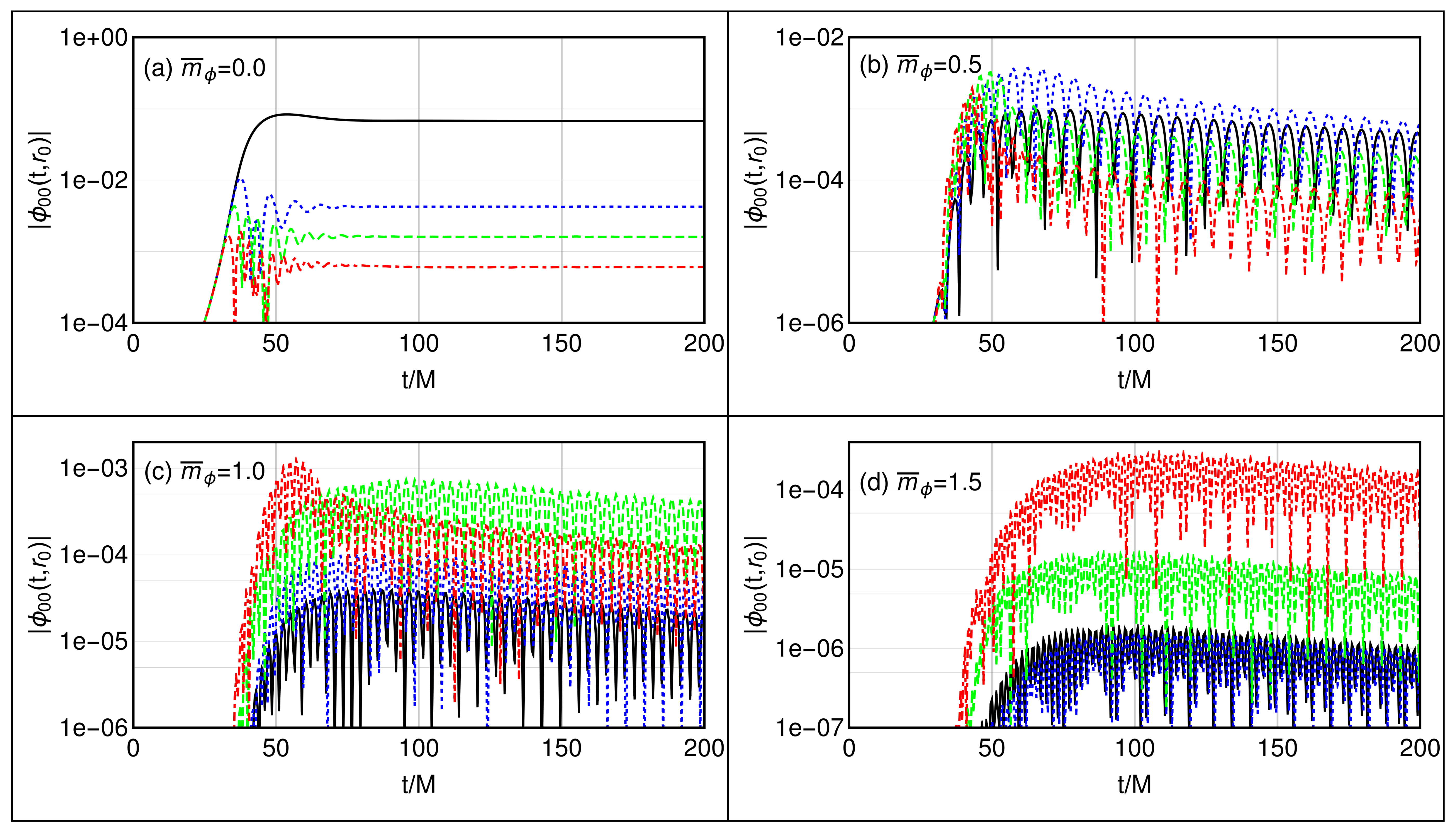}
	\caption{Values of the scalar mode $\phi_{00}(t,r)$ measured at $r=30M$ as the functions of time for each case listed in Table \ref{values_of_parameters}. Here, the black solid line, blue dotted line, green dashed line, and red dotdasshed line stand for the results with $\lambda=0$, $\lambda=10$, $\lambda=100$, and $\lambda=1000$, respectively.}
	\label{mp_sf_00}
\end{figure*}

We measure the values of $\phi_{00}(t,r)$ for all the cases listed in Table \ref{values_of_parameters} at a fixed radius $r=30M$ and show them as a function of time in Fig. \ref{mp_sf_00}. We observe that the values of the massless scalar field quickly relax to constants. This behavior is consistent with the results described by the subfigure (a2) in Fig. \ref{radial_profiles_of_sf}. The oscillate behavior and decay law of the massive scalar field can be extracted in terms of the behavior of $\phi_{00}(t,r)$ in Fig. \ref{mp_sf_00}. Here we do not compute them.

The results in Figs. \ref{radial_profiles_of_sf} and \ref{mp_sf_00} have shown that the GB term will lead to a nontrivial configuration for the scalar field, the dynamical formation of these nontrivial configurations will induce the energy flux of the scalar field. In order to analyze the dynamics of scalar field, we compute the energy flux of the scalar field based on the stress-energy tensor \eqref{tmunu_phi} as follows
\begin{equation}
\frac{dE^{\text{sf}}}{dt}=\lim_{r\to r_c}r^2\oint{T_{tr}\,d\Omega}.
\label{energy_fluxes}
\end{equation}
The spherical Schwarzchild background ensures that the following flux of the linear momentum
\begin{equation}
\frac{dP_i^{\text{sf}}}{dt}=\lim_{r\to r_c}r^2\oint{ T_{ir}\,d\Omega}
\label{rediated-momentum-sf}
\end{equation}
and flux of the angular momentum
\begin{equation}
\frac{dJ_z^{\text{sf}}}{dt}=\lim_{r\to r_c}r^2\oint{T_{\phi r}\,d\Omega}
\label{rediated-angular-momentum-sf}
\end{equation}
are zero. Therefore, we only compute the energy flux \eqref{energy_fluxes} of the scalar field. Here, we compute the above integrals by setting $r_c=30M$. One can derive the corresponding radiated energy of the scalar field in terms of the energy flux \eqref{energy_fluxes} as follows
\begin{equation}
E_{\text{rad}}(t)=\int_0^t\frac{dE^{\text{sf}}}{dt'} dt'.
\label{rediated_energy}
\end{equation}

\begin{figure*}[!htb]
	\includegraphics[width=0.98\linewidth]{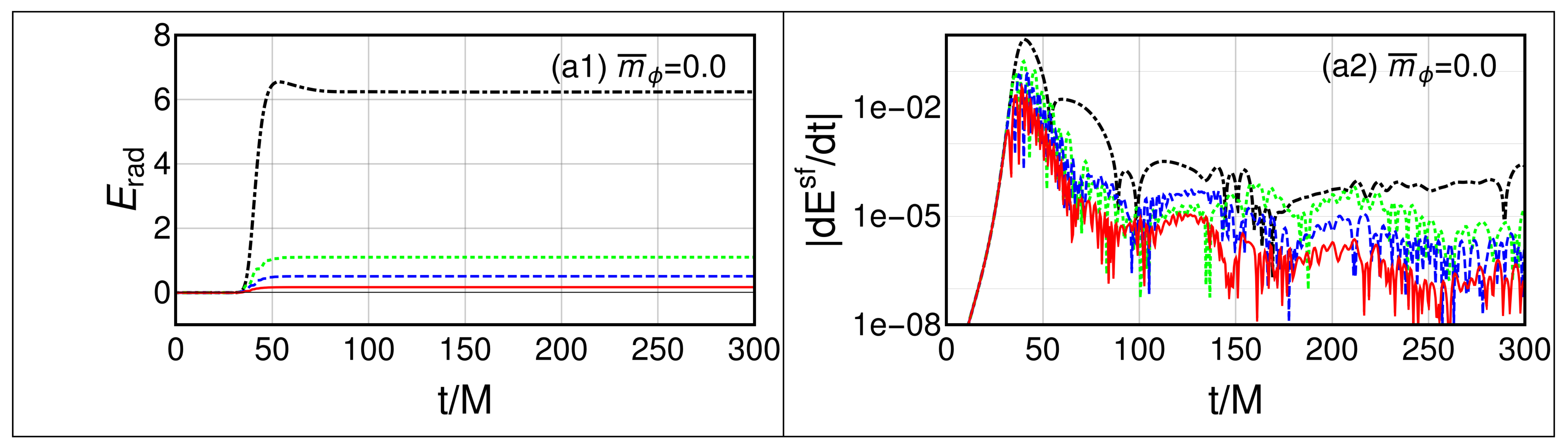}
	\includegraphics[width=0.98\linewidth]{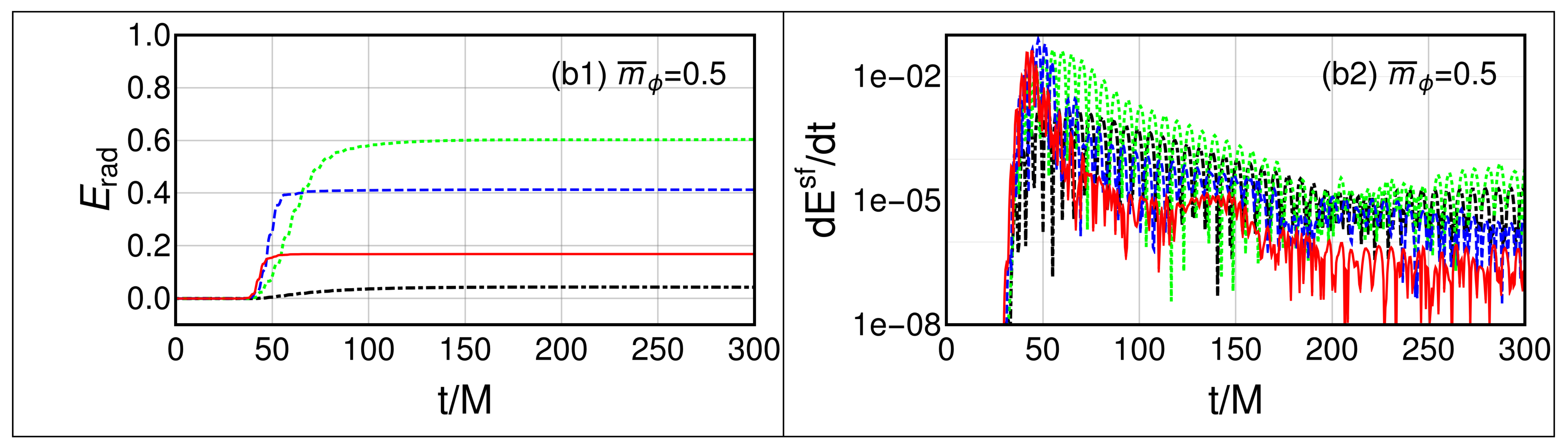}
	\includegraphics[width=0.98\linewidth]{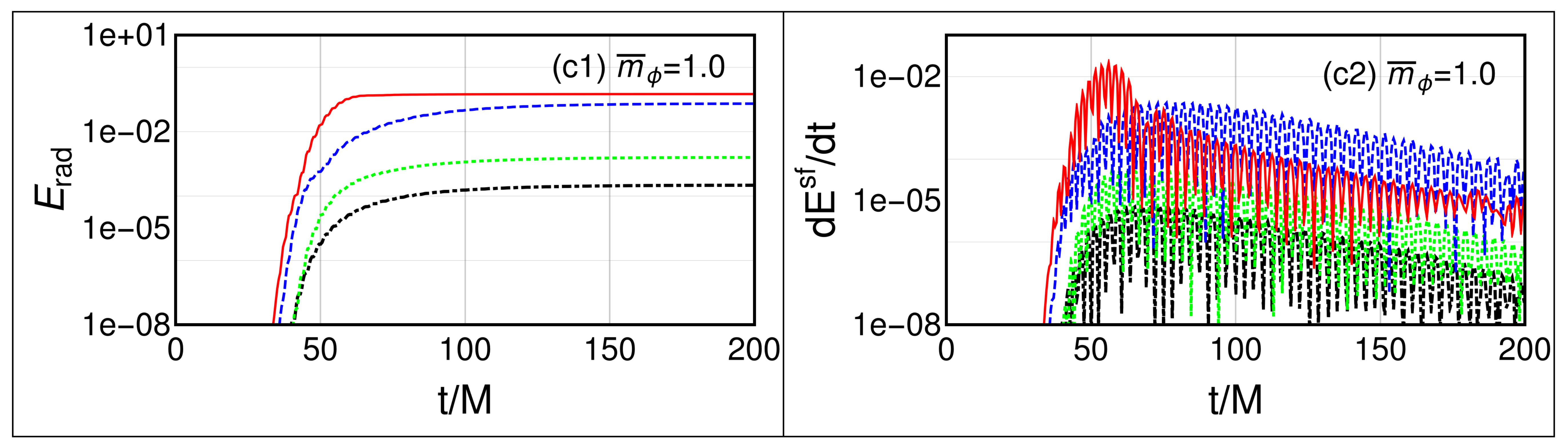}
	\includegraphics[width=0.98\linewidth]{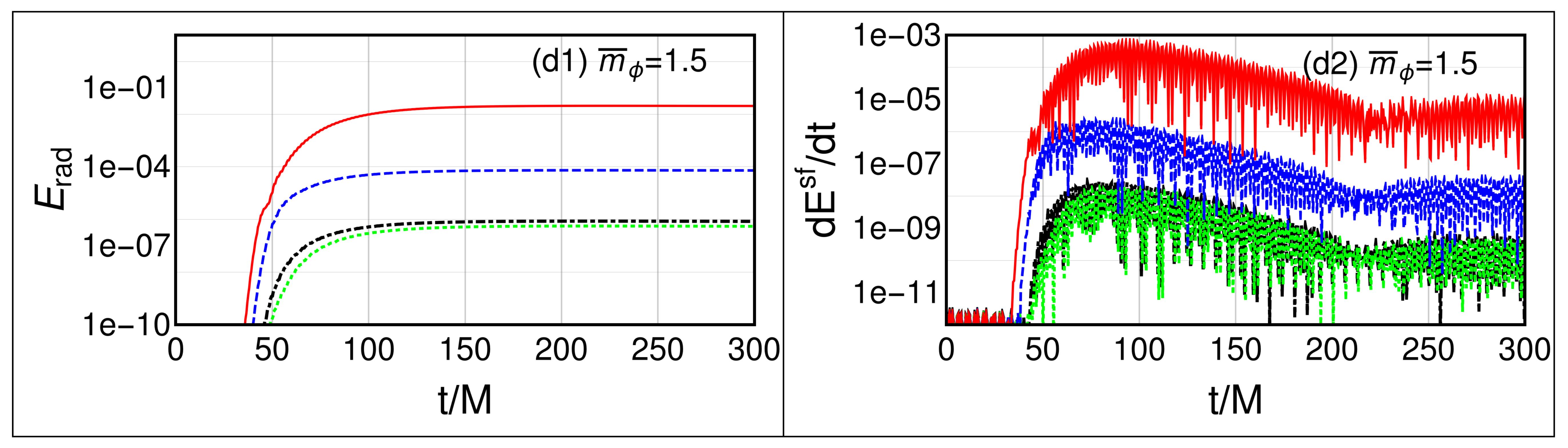}
	\caption{Radiated energies and powers $\frac{dE^{\text{sf}}}{dt}$  of scalar fields as the functions of time for all cases listed in Table \ref{values_of_parameters}. The black dotdashed line, green dotted line, blue dashed line, and red solid line stand for the results with $\lambda=0$, $\lambda=10$, $\lambda=100$, and $\lambda=1000$, respectively.}
	\label{sf_rad_dedt}
\end{figure*}

Figure \ref{sf_rad_dedt} shows the results of the energy flux $\frac{dE^{\text{sf}}}{dt}$ and the radiated energy $E_{\text{rad}}$ of the scalar field as functions of time. We observe that the radiated energy for the massless scalar field will reach their maxima quickly. While for the massive cases, the rate for the radiated energy approaches the maximum is less than the massless cases. For the massless cases (sf\_1\_1, sf\_1\_2, sf\_1\_3, sf\_1\_4), we observe that the radiated energies decrease with the coupling parameter $\lambda$. {While for the massive cases (sf\_2\_1, sf\_2\_2, sf\_2\_3, sf\_2\_4), (sf\_3\_1, sf\_3\_2, sf\_3\_3, sf\_3\_4), and (sf\_4\_1, sf\_4\_2, sf\_4\_3, sf\_4\_4), the corresponding radiated energies will increase or decrease with the coupling parameter $\lambda$. With the increase of the mass parameter of the scalar field, the dependence of the radiated energy on the coupling parameter $\lambda$ will be different.}

Besides the energy fluxes of the scalar field, we still compute the total energy of the scalar field by using the canonical energy-momentum tensor as follows
\begin{equation}
E_{\text{sf}}=\int{\alpha\sqrt{\gamma}\,\rho\,d^3x}.
\label{total_energy}
\end{equation}
Here we define the corresponding energy density of the scalar field in terms of the canonical energy-momentum \eqref{tmunu_phi}:
\begin{equation}
\rho=T_{\mu\nu}^{(\phi)}n^\mu n^\nu.
\end{equation}

\begin{figure*}[!htb]
	\includegraphics[width=\linewidth]{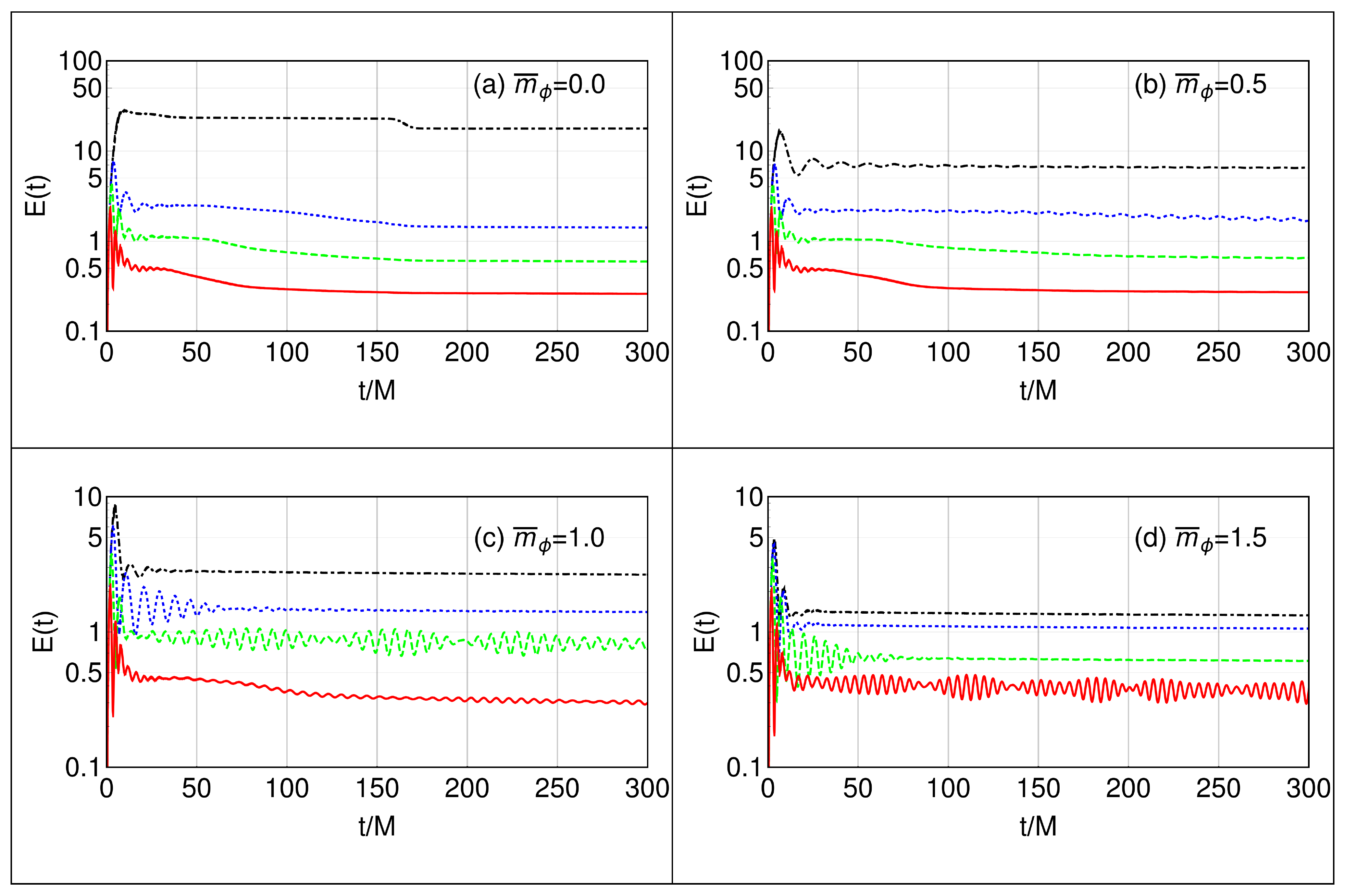}
	\caption{Energies of scalar fields as the functions of time for the cases listed in Table \ref{values_of_parameters}. The black dotdashed line, blue dotted line, green dashed line, and red solid line stand for the results with $\lambda=0$, $\lambda=10$, $\lambda=100$, and $\lambda=1000$, respectively.}
	\label{sf_energy}
\end{figure*}

We measure the total energy \eqref{total_energy} of {the scalar field} for the whole process of the evolution and obtain the total energy of the scalar field as a function of time. Due to the limit of the space for our simulation, we integrate the total energy \eqref{total_energy} from $r=0M$ to $r=120M$. {{Note that}, the profiles of the scalar {field} are nontrivial in the interior region of the apparent horizon and these parts actually do not contribute to the total energy of the scalar field. Therefore, we compute the total energy of the scalar field by excising the interior region of the apparent horizon.} The total energies of the scalar fields for all the cases listed in Table \ref{values_of_parameters} are given in Fig. \ref{sf_energy}. For all the cases, we observe that the energies increase rapidly at the initial moment, then decrease, and gradually tend to the constants. The oscillate behaviors of the scalar field are also reflected in the total energy of the scalar field and we also observe the oscillate behaviors for the total energies of the massive scalar fields.

\begin{figure*}[!htb]
	\includegraphics[width=\linewidth]{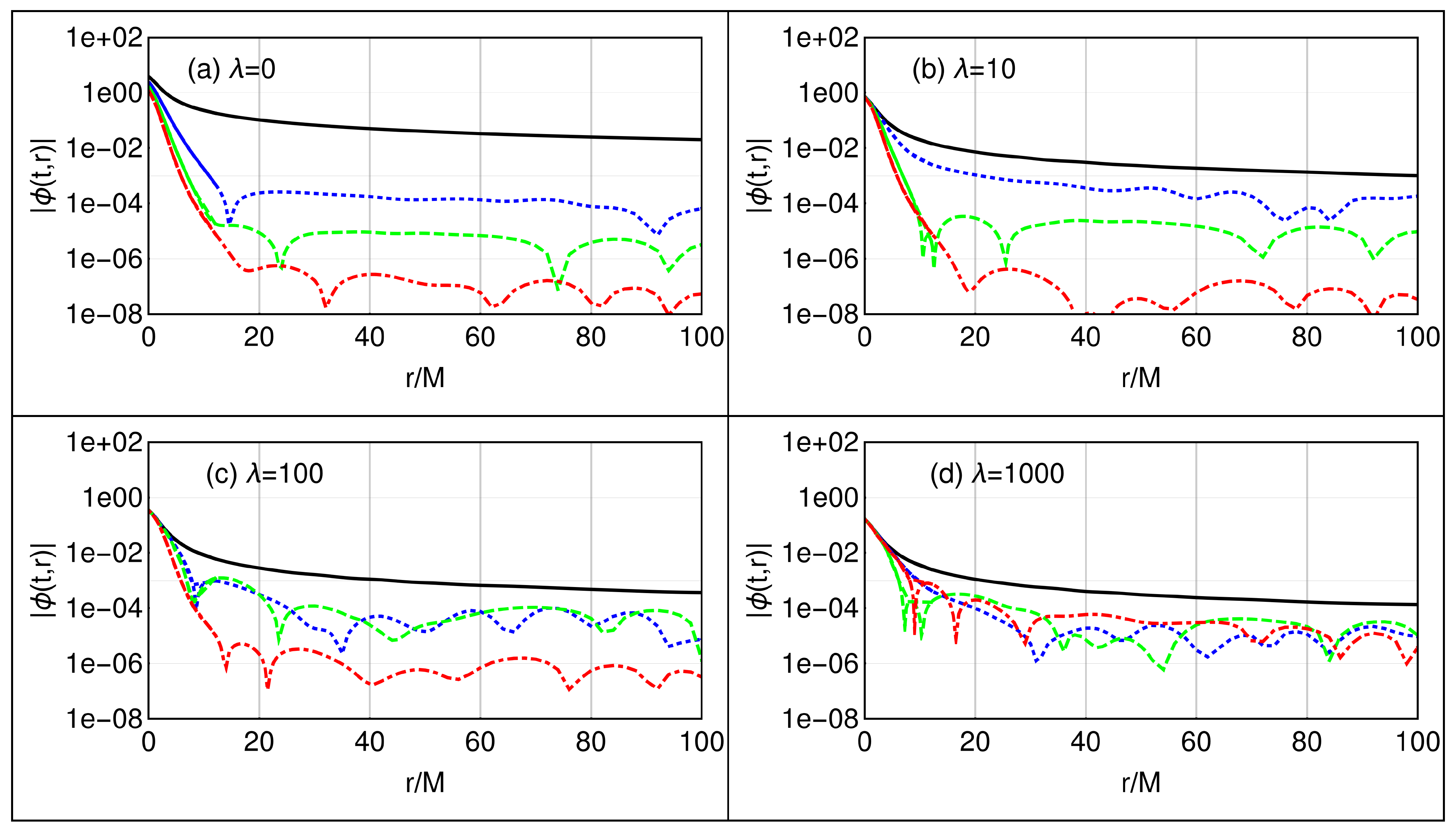}
	\caption{Comparisons of the radial profiles of the scalar field for the case with the fixed parameter $\lambda$ and varying mass parameters at late time $t=400M$. Here, the black solid line, blue dotted line, green dashed line, and red dotdashed line stand for the profiles of scalar fields with $\bar{m}_\phi=0$, $\bar{m}_\phi=0.5$, $\bar{m}_\phi=1.0$, and $\bar{m}_\phi=1.5$, respectively.}
	\label{sf_profile_massive}
\end{figure*}

{So far, we have obtained all the results of {the cases} listed in Table \ref{values_of_parameters}. We find that the introduction of {the mass term} and the quartic self-interaction \eqref{potentialofphi} {suppresses} the existence of the scalar {field}. For a fixed mass parameter, we find that the total energy decreases with the coupling parameter $\lambda$. For a fixed coupling parameter $\lambda$, {the total energy} of the scalar field decreases with the mass parameter $m_\phi$. Such behaviors are consistent with the results {found} in Refs. \cite{Doneva:2019vuh,Macedo:2019sem}.

In Ref.~\cite{Gregory:1992kr}, the authors studied the properties of {a black hole} in massive dilaton gravity. They showed that the mass term of the scalar field will suppress {the scalar field} at a length larger than the Compton wavelength $\lambda_\phi$. While at a length smaller than the Compton wavelength $\lambda_\phi$, the mass term behaves like the massless case. Thus, we expect that the scalar {field} with different mass parameters should have the similar {behavior} described in Ref. \cite{Gregory:1992kr}. Note that, we focus on the dynamics and final configurations of {the scalar field} around {the} Schwarzchild black hole in this paper. To validate if the scalar {field has} the behaviors described in Ref.~\cite{Gregory:1992kr}, we should consider the configurations of {the scalar field} at late time. Therefore, we compare them with a fixed $\lambda$ and varying $m_\phi$ in the near or far regions of {the} black hole. We give the corresponding comparisons in Fig.~\ref{sf_profile_massive}. We find that the profiles of {the scalar field} are almost same in the near region of {the} black hole. In the far region of {the} black hole, we find that the values of {the} scalar field decrease with the mass parameter. We still observe that the effects of {the} mass term on the scalar field also {depend} on the quartic interaction $\lambda \phi^4$. However, for all the cases listed in Table \ref*{values_of_parameters}, the suppression of {the} mass term on the configurations of {the} scalar field in the near or far region of {the} black hole are the almost the same although they are sourced by the GB term. Finally, we should note that we work perturbatively in the coupling constant and ignore the back-reaction of the scalar hair and the Gauss-Bonnet invariant.}

\section{Conclusions and outlook}\label{Conclusion}

In this paper, we investigated the dynamical formation of scalar hair with the self-interaction \eqref{potentialofphi} in the scalar GB gravity. By considering a linear coupling between the scalar field and GB term and ignoring the back-reaction of them, we evolved the scalar field sourced by a GB term in the background of Schwarzchild black hole described by GR. We evolved $16$ cases (see details in Table \ref{values_of_parameters}) in total with different parameters for the self-interaction \eqref{potentialofphi} and investigated how the dynamics of the scalar field depend on the self-interaction.

We obtained the configurations of the scalar field at different instances and found that all the scalar fields will relax to static configurations at late time. By projecting the scalar field with spherical harmonics, we found that the massless scalar field quickly relaxes to a static configuration and there is no oscillate behavior at late time. However, the massive scalar field has the oscillate behavior. We also computed the energy flux of the scalar field and derived the corresponding radiated energy. For the massless scalar field, we found that the radiated energy decreases with the coupling parameter $\lambda$. For the massive cases, the radiated energies will increase or decrease with the coupling parameter $\lambda$, and it depends on the mass parameter $m_\phi$. Finally, we computed the total energy of scalar field in a spherical space with a radius $r=120M$ in terms of the canonical energy-momentum tensor \eqref{tmunu_phi}. We found that the introduction of the self-interaction \eqref{potentialofphi} reduces the total energy of the scalar hairs at late time, and the corresponding total energy decreases with the mass parameter $m$ and the coupling parameter $\lambda$.

Our results show that how the dynamics of the scalar field depends on the self-interaction \eqref{potentialofphi} in the scalar GB gravity within a perturbative approach in the coupling constant. This will be helpful for understanding the dynamical scalarization of a black hole in the scalar GB gravity and the possible configuration of the scalar hair.

\section{Acknowledgments}

This work was supported in part by the National Key Research and Development Program of China (Grant No. 2020YFC2201503), the National Natural Science Foundation of China (Grants No. 12105126, No. 11875151, No. 12075103, and No. 12047501), the Fundamental Research Funds for the Central Universities (Grant No. lzujbky-2021-pd08), the China Postdoctoral Science Foundation (Grant No. 2021M701531), and the 111 Project (Grant No. B20063). Y.X. Liu was supported by Lanzhou City's scientific research funding subsidy to Lanzhou University.

\section*{Appendix: convergence test} 
\begin{figure*}[!htb]
	\includegraphics[width=0.98\textwidth]{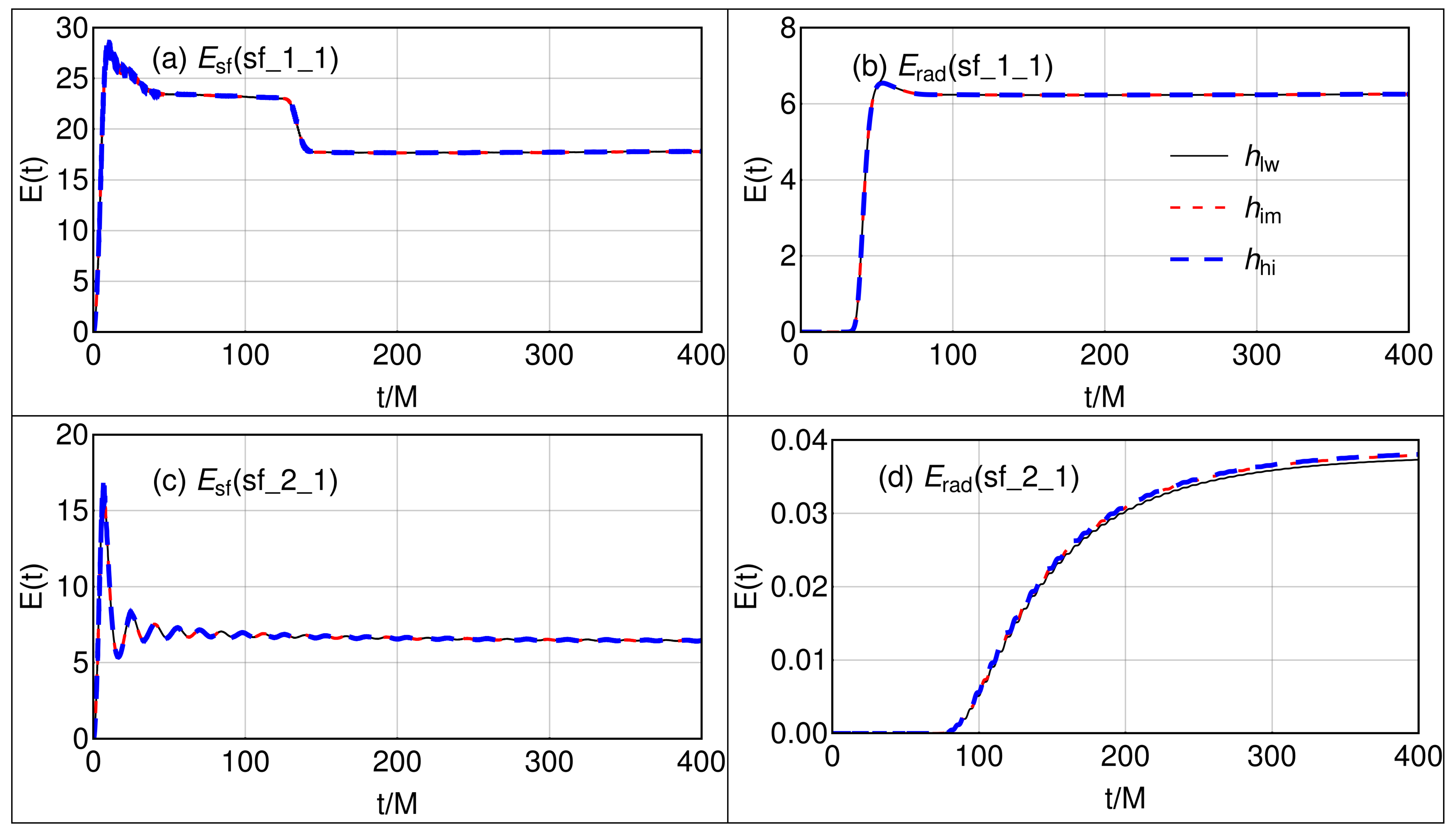}
	\vskip -0mm \caption{Comparison of the radiated energy and total energy for the cases (sf\_1\_1, sf\_2\_1) with three different resolutions. Here, the black solid line, the blue thick dashed line, and the red dashed line stand for the results with low, immediate, and high resolutions, respectively.}
	\label{convergencetest1}
\end{figure*}

\begin{figure*}[!htb]
	\includegraphics[width=0.98\textwidth]{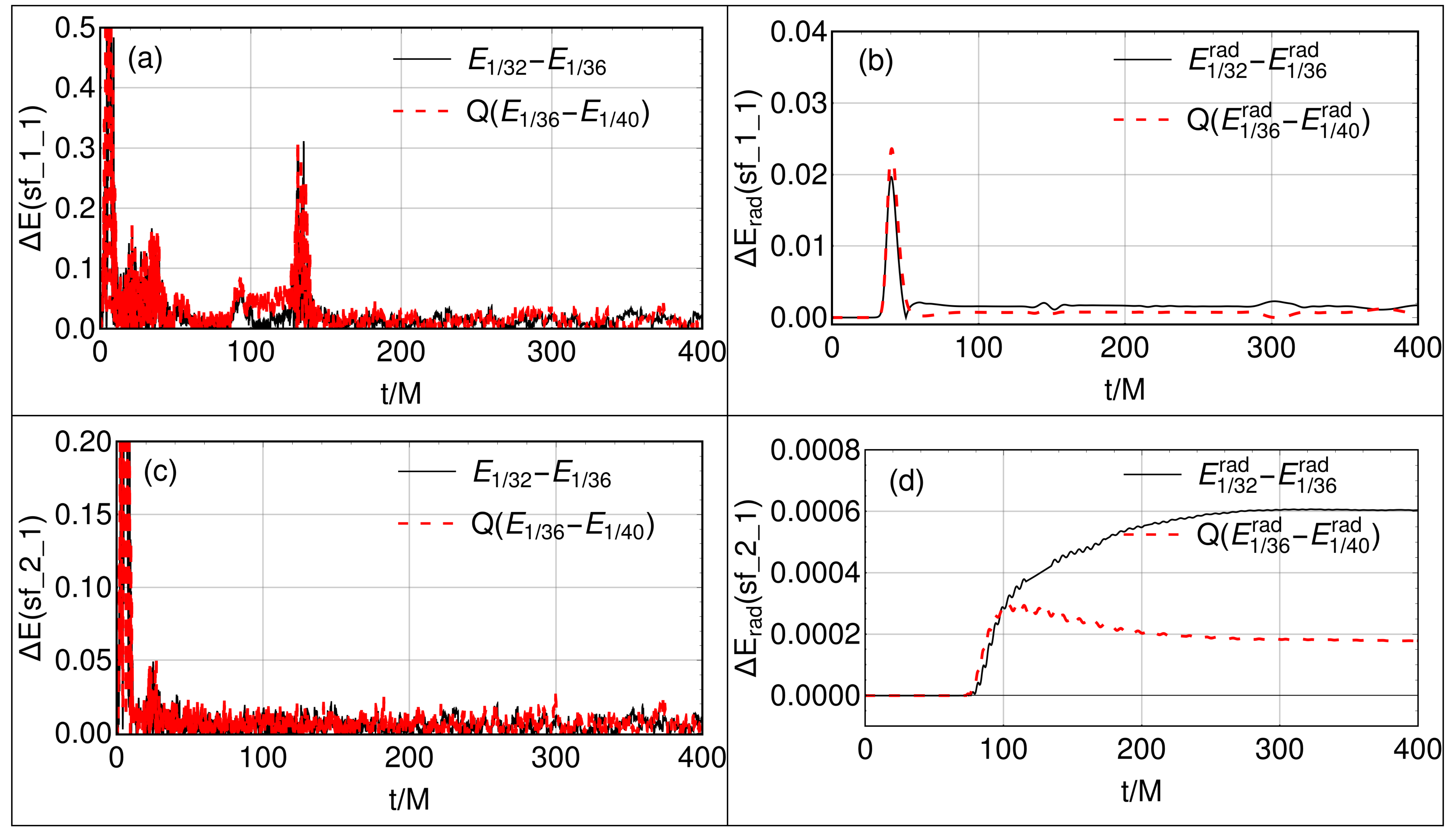}
	\vskip -0mm \caption{Convergence plots for the cases sf\_1\_1 and sf\_2\_1 with three different resolutions. Here, we present the differences in the coarse-medium-resolution and medium-high-resolution.}
	\label{convergencetest2}
\end{figure*}

{In this appendix, we will briefly discuss the numerical accuracy of our simulations. To do this, we evolve two cases (sf\_1\_1, sf\_1\_1) to test the convergence of our simulation with three resolutions $h_{\text{lw}}=M/32$, $h_{\text{im}}=M/36$, $h_{\text{hi}}=M/40$ and compare the corresponding total energy and radiated energy. The total energy {is computed} within the inner of a spherical surface with radius $r=120M$, and the radiated energy of the scalar field {is extracted} at the spherical surface with radius $r=30M$.
	
In general, the numerical solutions of a system converge according following rule
\begin{equation}
\phi_{h} - \phi\propto h^n,\label{convergencerule}
\end{equation}
where $h$ is the corresponding resolution and $n$ is the convergence order. Therefore, one can obtain the relation as follows
\begin{eqnarray}
\frac{\phi_{h_{\text{lw}}}-\phi_{h_{\text{im}}}}{\phi_{h_{\text{im}}}-\phi_{h_{\text{hi}}}}=\frac{h_{\text{lw}}^n-h_{\text{im}}^n}{h_{\text{im}}^n-h_{\text{hi}}^n}=Q.
\end{eqnarray}
We give the corresponding convergence plots in Figs. \ref{convergencetest1} and \ref{convergencetest2} based on the total and radiated energies of the scalar field.  We find second-order convergence as indicated by the factor $Q\simeq 1.40$. Here the convergence factor $Q$ is
\begin{equation}
Q=\frac{(1/32)^n-(1/36)^n}{(1/36)^n-(1/40)^n}.
\end{equation}
$Q\simeq 1.40$ when $n=2$ means the convergence order is $2$. Note that although we have used the fourth-order derivative stencils in our code, the interpolation schemes at the refinement boundaries also affect the convergences, which induces {that} the convergence order of our code is about $2$.  For the two cases sf\_1\_1 and sf\_2\_1, we observe that the radiated energy \eqref{rediated_energy} and the total energy \eqref{total_energy} with different resolutions exhibit a maximal numerical error about $2\%$ at late time, such error means our resolution is enough.}

\end{document}